\documentstyle[aps,amssymb,multicol]{revtex}
\tightenlines
\begin{document}

\def\CC{{\rm\kern.24em \vrule width.04em height1.46ex depth-.07ex
\kern-.30em C}}
\def\P{{\rm I\kern-.25em P}}
\def\RR{{\rm
         \vrule width.04em height1.58ex depth-.0ex
         \kern-.04em R}}
\def\id{{\rm 1\kern-.22em l}}
\def\up{\uparrow}
\def\dwn{\downarrow}
\newcommand{\ket}[1]{\left | #1 \right\rangle}

\title{Integrable model for interacting electrons in metallic grains}
	
\author{Luigi Amico, Antonio Di Lorenzo, 
and Andreas Osterloh}
\address{Dipartimento di Metodologie Fisiche e Chimiche (DMFCI), 
	Universit\`a di Catania, viale A. Doria 6, I-95125 Catania, Italy}
\address{Istituto Nazionale per la Fisica della Materia, Unit\`a  di Catania, Italy}

\maketitle

\begin{abstract}
We find  an integrable generalization of the BCS  model with
non-uniform  Coulomb and pairing interaction. 
The  Hamiltonian  is integrable by construction since it is
a functional of commuting operators; these operators, which therefore 
are constants of motion of the model, contain the anisotropic Gaudin 
Hamiltonians.
The exact solution is obtained diagonalizing them
by means of  Bethe Ansatz.  
Uniform pairing and Coulomb interaction 
are obtained as  the ``isotropic limit'' of the Gaudin Hamiltonians. 
We discuss possible applications of this model 
to  a single grain 
and to a system of few interacting grains.
\end{abstract}
{\it PACS} N. 74.20.Fk, 02.30.Ik

\begin{multicols}{2}
\narrowtext
{\it Introduction and summary of the results.} Progress in nanotechnology 
has opened up theoretical 
investigations on the behavior of disordered interacting 
systems of small size~\cite{DOTS}. Recently,  the $I$--$V$ 
characteristic measurements of Ralph, Black and Tinkham\cite{BRT} 
on small $Al$ dots 
 stimulated the theoretical debate on how to 
characterize the physical properties of small metallic grains, 
such as superconductivity and ferromagnetism~\cite{ALTSHULER,MASTELLONE}.  
Due to the chaoticity of the single particle 
wave functions~\cite{DOTS,ALTSHULER}, 
the Hamiltonian of these systems reads
\begin{eqnarray}
H_{grain} &  = &\sum_{i} 
\varepsilon_{i} n_{i \sigma}
-  g \sum_{i, j}\, c_{i\up}^\dagger c_{i \dwn}^\dagger 
c_{j \dwn} c_{j \up} + 
U {\left(\sum_{{j}}
n_{j \sigma}\right)^2}  \nonumber \\
\mbox{ } &-& J\left(\sum_{j} c^\dagger_{j\sigma}
\vec{S}_{\sigma\sigma'} 
c_{j\sigma'}\right)^2 +\;{\cal O}({\delta E^2}/{E_T}). 
\label{universal}
\end{eqnarray}
(Here and in the following, sums over spins $\sigma, \sigma'$ are implied).
The quantum numbers $i$, $\sigma$ 
label a shell of doubly 
degenerate single particle energy levels
with energy $\epsilon_i$ and annihilation operator 
$c_{i\sigma}$; $n_{i \sigma}:=c_{i \sigma}^\dagger c_{i \sigma}$; $S^a$, 
$a=x,y,z$, are $2 \times 2 $ spin matrices; $\delta E$ is the average level spacing, 
and $E_T$ the Thouless energy.
The  {\it universal} part of the Hamiltonian~(\ref{universal}) (namely the 
first four terms) describes 
the pairing attraction, the electrostatic interaction 
and the ferromagnetic instability, respectively. 
\\
The superconducting fluctuations~\cite{MASTELLONE}
can be taken into 
account  by employing  the  BCS model~\cite{BCS,IACHELLO} 
(namely taking the first two terms 
in Eq.~(\ref{universal})).  
Richardson and Sherman (RS)~\cite{RICHARDSON} 
constructed  the 
exact solution of the BCS model 
by a procedure close in spirit to the coordinate Bethe 
Ansatz (BA). 
The  knowledge of the exact eigenstates and eigenvalues of the BCS model has 
been  crucial to establish physically relevant  
observables~\cite{FALCI}.   
The integrability of the model has been 
proved~\cite{SKLYANIN-GAUDIN,CAMBIAGGIO} to be
deeply related to the integrability 
of the isotropic Gaudin magnet~\cite{GAUDIN}: the BCS model can be expressed 
as a certain 
combination (see Eq.~(\ref{crs}) below) of its integrals of motion, which 
contain Gaudin Hamiltonians. Relations with 
conformal  field theory and disordered vertex models were investigated 
in Refs.~\cite{SIERRA,AMICO}. 
\\
Many properties of  metallic grains in a normal state (negligible 
superconducting fluctuations)
can be described by the orthodox 
model~\cite{DOTS,AVERIN} 
(i.e. taking the first and the third term 
of the Hamiltonian~(\ref{universal})). 
This arises by assuming uniform Coulomb interaction. 
\\
Magnetic phenomena like the mesoscopic Stoner 
instability~\cite{ALTSHULER} can be 
studied by means of the exchange  contribution to the Hamiltonian (the fourth 
term in Eq.~(\ref{universal})). \\ 
The terms proportional to $\delta E^2/E_T$ 
correspond to non uniform  
Coulomb interaction\cite{AGAM1}. Although they lose  
importance with  the increasing conductance of the system, 
these corrections gain physical relevance due to the 
typically low relaxation rate of the excitations in a small dot. In fact  
the corrections to the orthodox model 
 induce ``fluctuations'' which can 
explain how non-equilibrium 
excitations decay in the dot~\cite{AGAM,SIVAN}. This results in   
the formation of clusters of resonance peaks in the tunneling 
spectroscopy experiments~\cite{BRT}. 

In this paper we present an integrable generalization of the BCS 
Hamiltonian with non-uniform pairing coupling $g_{ij}$ and solve it
exactly. Besides the non-uniform pairing, the Hamiltonian
contains a non-uniform Coulomb interaction $U_{ij}$; 
$g_{ij}$ and $U_{ij}$ are fixed according to 
Eqs.~(\ref{coupling-constants}).
We shall see that the inclusion of certain ${\cal O}(\delta E^2/E_T)$ 
terms leads to our integrable  model. 
The integrable Hamiltonian we solve is
\begin{eqnarray}
H &= &\sum_{i}
\varepsilon_{i} n_{i \sigma}
  -   \sum_{i, j} g_{ij}\, c_{i\up}^\dagger c_{i \dwn}^\dagger 
c_{j \dwn} c_{j \up} \nonumber \\
\mbox{} &+&{\sum_{{i, j}}}   U_{ij} 
n_{i \sigma} n_{j \sigma'} - 
J \left(\sum_{j} c^\dagger_{j\sigma} \vec{S}_{\sigma\sigma'} 
c_{j\sigma'}\right)^2 \quad,
\label{our-model}
\end{eqnarray} 
where the couplings are
\begin{eqnarray}
\label{coupling-constants}
&&\left\{ 
\begin{array}l
g_{ij} = - q K{\left(\varepsilon_{i}-\varepsilon_{j}\right)}
/{ \sinh{q (u_{i}-u_{j})} }\;, \\
\quad \\
4 U_{ij} = A + q K (\varepsilon_{i} - \varepsilon_{j}) \coth{q (u_{i}-u_{j})},
\end{array} 
\right. i\neq j \nonumber \\
&&\left. 
\begin{array}l
\quad g_{jj} = - \beta_{j}\;, \qquad   
\quad 4 U_{jj} = A + \beta_{j}\; , 
\end{array}
\right. 
\end{eqnarray} 
where $2\beta_j= - qK \sum_{i\neq j} 
(\varepsilon_i-\varepsilon_j) \coth{q (u_{i}-u_{j})} 
+ C$.\cite{RICHARDSON-PRIV} 
For generic choices of $\beta_j$, the single particle energies 
$\varepsilon_j$ must be shifted by
$2\beta_j + 4 \sum_{i\neq j} U_{i,j}$ in order to have integrability. 
The parameters $A$, $K$, and $C$ are arbitrary real constants,
while $q$ can be real or imaginary. 
The BCS Hamiltonian, including a tunable capacitive coupling 
can be obtained from (\ref{our-model}) in the {\it isotropic limit} $q\to 0$. 
Non-uniform coupling constants are obtained for generic $q$, and $u_j$ being 
monotonic functions of 
$\varepsilon_j$. For real $q$, the arising $g_{ij}$ can be made nearly uniform 
for levels within an energy cutoff $E_D$, 
and exponentially suppressed otherwise; 
correspondingly, $U_{ij}$ can be made nearly uniform (as specified in 
Eqs.(\ref{single}) below). 
\\
The proof of the integrability of the Hamiltonian~(\ref{our-model}) 
proceeds  along  the two following steps.
{\it i)}~First we note the  factorization~\cite{NOTE2} of 
the  eigenstates of the Hamiltonian~(\ref{our-model}):
$\ket{\Psi}=\ket{\Psi_N}\otimes \ket{\Phi_M}$ with eigenvalue $E=E_N+E_M$; 
where $\ket{\Psi_N}$ 
is the eigenstate of the Hamiltonian $H_N$ projected on the subspace with $N$ 
time-reversed pairs; 
$\ket{\Phi_M}$ is the Fock state projected  on the blocked 
$M$ singly occupied levels. The  
solution of the corresponding Hamiltonian $H_M$ 
is easily obtained~\cite{NOTE3} as: $H_M \ket{\Phi_M}=
[\sum_i \varepsilon_i+\sum_{ij}U_{ij}-J S(S+1)] \ket{\Phi_M}$.  
{\it ii)}~Then Hamiltonian~(\ref{our-model}) 
is integrable if and only if 
$H_N$ is integrable. 
The Hamiltonian $H_N$ is obtained by inverting the procedure  presented 
in Ref.~\cite{CAMBIAGGIO}: First, we  modify the 
constants of motion (of the BCS model) 
to commuting operators containing the anisotropic Gaudin models 
(the isotropic ones being considered in \cite{CAMBIAGGIO});
then we define the Hamiltonian in terms of these operators
($H_N$ is therefore integrable by construction). We discuss some choices of 
$\{u_j\}$, $K$, and $A$ leading to physically relevant Hamiltonians. 
The exact solution of $H_N$ 
is found by diagonalizing the  integrals of motion through BA~\cite{GAUDIN}.
The exact eigenfunctions and eigenvalues $\Psi_N$, 
$E_N$ are 
\begin{eqnarray}
\Psi_N &=& \prod_{\alpha=1}^{N}\sum_{j=1}^{\Omega}  
\frac{q c^{\dagger}_{j\up} 
c^{\dagger}_{j\dwn}}{\sinh q(\omega_\alpha - u_j)} \ket{0} 
\label{vector} \\
E_N &=&  qK \sum_{j=1}^\Omega  \sum_{\alpha=1}^N
\varepsilon_{j} \coth q ( \omega_\alpha - u_j )
+ A N^2 \quad ;
\label{eigen}
\end{eqnarray}
$\ket{0}$ is the electronic vacuum state and $\Omega$ is the number of levels. 
The quantities $\omega_\alpha$ fulfill the  equations
\begin{eqnarray}
\frac{2}{K}&-&\sum_{l=1}^\Omega q\;{\rm coth}\;{q( \omega_\alpha - u_l )} 
\nonumber \\
	\mbox{} &+& 2\sum_{\beta=1 \atop \beta\neq\alpha}^N   
	q\; {\rm coth}\; {q ( \omega_\alpha - \omega_\beta )} = 0 \;, 
\alpha=1,\dots,N
\label{generalized-richardson}
\end{eqnarray}

Our results can be applied to describe a system of $\cal N$ grains, 
since their Hamiltonian can be written (after a suitable 
relabeling of the levels) in the form (\ref{our-model}). For distinct grains 
$g_{jk}$ describe Cooper pair tunneling, and $U_{jk}$ the inter-grain 
Coulomb interaction. We require $g_{ij}$ 
to decay both with inter-grain distance and level separation. 
This can be fulfilled with  $u_{j}$ fixed by 
Eqs.(\ref{multi}),~(\ref{multi1}).

The present paper is laid out as follows. First we discuss the integrability 
of the model $H_N$; then its  exact solution is presented. This 
will complete the study of the integrability of the 
Hamiltonian~(\ref{our-model}).
Finally, we will explain how our model  can be applied 
to describe single as well many interacting grains. 

\vspace{-0.2cm}
\bigskip  

{\it Integrability.} 
The BCS Hamiltonian can be written in terms of
the spin-$1/2$ realization of $su(2)$: \\
$
H_{BCS}=\displaystyle\sum\limits_{j} 2 \varepsilon_{j} S^{z}_{j}
- g \displaystyle\sum\limits_{j,k} S^{+}_{j} S^{-}_{k}, 
$
where 
\begin{eqnarray}
S^-_j &:=& c_{j\dwn} c_{j\up}, \;\; S^+_j:=(S^-_j)^\dagger= 
c^\dagger_{j\up} c^\dagger_{j\dwn} , \;\; \nonumber \\ 
S^z_j &:=& \frac{1}{2}(c^\dagger_{j\up} c_{j\up}+ c^\dagger_{j\dwn} c_{j\dwn}-1), 
\end{eqnarray}
obeying
$
\left [ S^z_{j}, S^\pm_{k} \right ] = 
\pm  \delta_{jk} S^{\pm}_{k}$, 
$ \left [ S^+_{j}, S^-_{k} \right ] = 2 \delta_{jk} S^{z}_{k}$.
Its constants of motion are written in terms of isotropic 
Gaudin Hamiltonians $\tilde{\Xi}_j$
\begin{equation}
\tilde{\tau}_{j}= S^{z}_{j} - g\; \tilde{\Xi}_j ;\quad 
\tilde{\Xi}_j= \sum_{{k=1}\atop{k\neq j}}^{\Omega} 
\frac{{\bf S}_{j}\cdot {\bf S}_{k}}{\varepsilon_j -\varepsilon_k}.
\label{isotau}
\end{equation}
The $\tilde{\tau}_j$ mutually commute and we have
$[\tilde{\tau}_j \, ,\, \tilde{\tau}_k ] = 
[H\, , \, \tilde{\tau}_j] = 0$ for all 
$i,j\in\{1,\dots,\Omega \}$, 
because the BCS Hamiltonian can be written in terms of 
the $\tilde{\tau}_j$ only:
\begin{equation}
\label{crs}
H_{BCS} = \sum_{j} 2 \varepsilon_{j} \tilde{\tau}_{j} 
+ g \sum_{j,k} \tilde{\tau}_{j}\tilde{\tau}_{k} .
\end{equation}
Our approach is now to modify the integrals of motion (\ref{isotau}) and  then 
to construct an integrable BCS-like model 
(which turns out to be characterized by a non-uniform pairing) following 
formula (\ref{crs}):
\begin{equation}
H_N := \sum_{j} 2 \varepsilon_{j} \tau_{j} 
+ A \sum_{j,k} \tau_{j}\tau_{k} + const.
\label{modified-crs}
\end{equation}
The ansatz for the modified integrals $\tau_j$ is
\begin{equation}
\tau_{j} = S^{z}_{j} + \Xi_j ;\quad 
\Xi_j= \sum_{{k=1}\atop{k\neq j}}^{\Omega} w_{jk}^{\alpha}
S_{j}^{\alpha} S_{k}^{\alpha}.
\label{constants}
\end{equation}
where the operators $\Xi_j$ are anisotropic 
Gaudin Hamiltonians (the isotropic case corresponding to $w_{ij}^{x}=
w_{ij}^{y}=w_{ij}^{z}$). 
These operators mutually commute if
\begin{eqnarray}
w_{ij}^{\alpha}w_{jk}^{\gamma}+w_{ji}^{\beta}w_{ik}^{\gamma}
&=& w_{ik}^{\alpha}w_{jk}^{\beta}, 
\label{comm1}\\
w_{ij}^{x} &=& - w_{ji}^{y}, 
\label{comm2}
\end{eqnarray}
where (\ref{comm1}) emerges from imposing $[\Xi_i,\Xi_j]=0$\cite{GAUDIN}.
The other condition arises from
$[S^{z}_{i}  ,\Xi_j ] + [\Xi_i, S^{z}_{j}  ]=0$.
We furthermore postulate particle number conservation,
which in the spin picture means 
$[\sum\limits_{i=1}^{\Omega} S^{z}_{i}  ,\Xi_j ] = 0$ 
for all $j\in\{1,\dots,\Omega\}$, leading to another condition
\begin{equation}
w_{ij}^{x} = w_{ij}^{y} \stackrel{{\rm Eq.}(\ref{comm2})}{=} 
-w_{ji}^{x} =: w_{ij} = - w_{ji}.
\end{equation}
The last equation reduces the anisotropy to the XXZ-type 
and Eqs.(\ref{comm1}) finally become
\begin{equation}
\label{Comm1}
w_{ij} v_{jk} + w_{ji} v_{ik}
= w_{ik} w_{jk} \;, \quad 
v_{ij}:= w_{ij}^{z}.
\end{equation}
The solution of Eqs.~(\ref{Comm1}) (see Ref.\cite{GAUDIN}) is 
\begin{eqnarray}
v_{jk} = qK \;\mbox{coth}\;{q(u_{j}-u_{k})},
\nonumber \\
w_{jk} =  \frac{qK}{\sinh{q(u_{j}-u_{k})}},\label{w-solution}
\end{eqnarray}
where $ u_{j}$ are arbitrary complex parameters
such that $v_{jk}, w_{jk}$ are real.
The transition from hyperbolic  to trigonometric functions in the
solution (\ref{w-solution}), is gained
through the choice $q = i$, with real $K$, $u_{j}$. 
The cubic and quartic terms in $S^{\alpha}_{j}$ (obtained from 
formula~(\ref{modified-crs})) vanish for the antisymmetry of $v_{jk}$.
We finally obtain:
\begin{equation}
H_N=\sum_{j} 2 \varepsilon_{j} S_{j}^{z} -
\sum_{j,k} g_{jk} S_{j}^{+} S_{k}^{-} +
4 \sum_{j,k} U_{jk} S_{j}^{z} S_{k}^{z},
\label{ourclass}
\end{equation}
where the couplings are given in Eqs.~(\ref{coupling-constants}).
Up to a constant, the Hamiltonian~(\ref{our-model}) (projected on doubly occupied states)
is recovered writing back the spin operators in terms of creation and 
annihilation operators. 

\vspace{-0.2cm}
\bigskip

{\it Exact solution.}
The exact solution of the anisotropic Gaudin model for  
$w_{ij}$ and $v_{ij}$ fixed by Eqs.~(\ref{w-solution})
was obtained in Ref.~\cite{GAUDIN}. The same  procedure can be applied 
to diagonalize $\tau_j$.
The eigenfunctions of $\tau_j$ 
defined in Eq.(\ref{constants}) are written in the form
\begin{eqnarray}
\ket{\Psi_j} &=& \mathop{{\sum}'}_{j_1\leq\dots\leq j_M}
	c(j_1,\dots,j_M) S^+_{j_1}\dots S^+_{j_M}\ket{0} \\
\mbox{}&+&\mathop{{\sum}'}_{j_1\leq\dots\leq j_{M-1}}
	e(j_1,\dots,j_{M-1}) S^+_{j_1}\dots S^+_{j_{M-1}}S^+_j\ket{0}. 
\nonumber
\end{eqnarray}
The vacuum $\ket{0}$ corresponds to $\ket{\downarrow,\dots,\downarrow}$ 
; the prime on the sums means the indices run in the range 
$\{1,\dots,\Omega\}\slash \{j\}$. 
Imposing that $\ket{\Psi_j}$ is an eigenstate of 
$\tau_j$  we find a  set of equations which $c(\{j_i\})$ and 
$e(\{j_i\})$ must fulfill.
For a suitable change of variables we find that these conditions are 
transformed  in Eqs.(\ref{generalized-richardson}).
The quantities $\tau_j$ have the following eigenvalues:
\begin{eqnarray}
&&\tau_j \ket{\Psi_j} =\frac{1}{2}\left (h_j-1\right ) \ket{\Psi_j} \\ 
&&\frac{1}{K} h_j= \frac{1}{2}{\sum_{l=1}^\Omega }' 
q\;{\rm coth}\;{q( u_j- u_l )} 
	- \sum_{\alpha=1}^N  
	q\; {\rm coth}\; {q ( u_j - \omega_\alpha )} \; . \nonumber 
\end{eqnarray}
The  parameters $\omega_\alpha$ are determined by 
Eq.(\ref{generalized-richardson}). 
The  eigenvalues of $H_N$ immediately follow from 
formula~(\ref{modified-crs}). Together with the eigenfunctions they are 
given in Eqs.(\ref{vector}),(\ref{eigen}). 

\vspace{-0.2cm}
\bigskip

{\it Single grain.}
We discuss how our results can be applied to describe the physics of a
single grain. 
\\
The isotropic limit $q\to 0$ of Eqs.(\ref{coupling-constants}) 
gives the BCS Hamiltonian plus a tunable capacitive coupling $A+g$, with 
$K = {g}/{E_D}$, $\beta_i = -g$, 
$u_{j} = -{\varepsilon_j}/{(E_{D}~\Theta(|\varepsilon_j-E_F|-E_D))}$, 
where $E_F$ is the Fermi level, and $\Theta$ is the Heaviside function 
($\Theta(x)=1$ if $x<0$, $\Theta(x)=0$ if $x>0$), setting  sharp cutoffs  
at the Debye energy\cite{NOTE4}; 
the diagonal elements  $U_{jj}$ 
and $g_{jj}$ can be independently set to arbitrary values 
(since they would renormalize $\varepsilon_j$).
Choosing $A=-g$ gives the ``pure'' BCS model. 
In this limit, the eigenstates and eigenvalues 
Eqs.(\ref{eigen}),(\ref{vector}) 
coincide with those of the BCS model and
Eqs. (\ref{generalized-richardson}) reduces to the RS equations~\cite{RICHARDSON}.

We now discuss the case corresponding to $q=1$: 
\begin{eqnarray}
 K = {g}/{E_D}\, ,\; \beta_i &=& -g \, ,\;  A\gg (g/E_D) \max_{j,k}
	\{\varepsilon_j-\varepsilon_k\}\nonumber \\
u_{j} &=& -\varepsilon_j/E_{D}\,.
\label{single}
\end{eqnarray}
We can identify three regimes depending on the value of $E_D$:
{\it i)} $E_D < \delta E$,  $g_{ij}$ is nearly zero, 
while $U_{ij} \simeq A - g$; {\it ii)} $E_D \sim \delta E$,  
the pairing interaction decays on the  scale  $E_D\sim \delta E$, 
while  $U_{ij}$ is slowly 
modulated by the energy separation; 
{\it iii)} $E_D > \max_{i,j} (\varepsilon_i-\varepsilon_j)$ both $g_{ij}$ and 
$U_{ij}$ are nearly uniform. 
 
\vspace{-0.2cm}
\bigskip

{\it Application to many interacting grains.} 
We now discuss  applications of the model~(\ref{our-model}) to 
interacting dots.
\\
The Hamiltonian (\ref{our-model}) can be reinterpreted as follows: 
the set $I=\{1,\dots, \Omega\}$ can be split into the (disjoint) sets 
$I_a$, $a=1 \dots {\cal N}$ containing the levels of the $a$-th grain:  
$I = \bigcup_a I_a$; 
$\Omega=\sum_{a=1}^{\cal N} \Omega_a$, where $\Omega_a=|I_a|$. 
Thus the Hamiltonian $H_N$ is equivalent to the following one:
\begin{eqnarray}
\label{multi-grain}
H_{\cal N}  &=& \sum_{a=1}^{\cal N} \sum_{i_a} 
\varepsilon^{(a)}_{i_a} c_{a,i_a \sigma}^\dagger c_{a,i_a \sigma} \\
\mbox{}&-& \sum_{a,b=1}^{\cal N}\sum_{i_a,j_b}   
g^{(a,b)}_{i_a j_b}\, 
c_{a,i_a\up}^\dagger c_{a, i_a \dwn}^\dagger 
c_{b,j_b \dwn} c_{b,j_b \up} \nonumber \\
\mbox{} &+&\sum_{a,b=1}^{\cal N}\sum_{i_a,j_b}
U^{(a,b)}_{i_aj_b} 
n_{a,i_a \sigma} n_{b,j_b \sigma'}\quad, \nonumber 
\end{eqnarray} 
where $i_{a}=1,\dots\Omega_a$ label the elements of $I_{a}$ 
and $c_{a,i_a \sigma}$ annihilates an electron with spin $\sigma$ in 
the $i_a$-th level of 
$a$-th grain. For $a\neq b$, $g^{(a,b)}$ describe 
tunneling of Cooper pairs; in terms of~(\ref{our-model}), 
$g^{(a,b)}_{i_a j_b} = g_{ij}$, where $i$ is the $i_a$-th element of $I_a$, 
and $j$ the $j_b$-th element of $I_b$; $U^{(a,b)}$ 
describe a Coulomb-like coupling between grains $a$ and $b$, and is 
written in terms of $U_{ij}$ analogously to $g^{(a,b)}$.
Couplings $g^{(a,a)}$ and $U^{(a,a)}$ describe pairing and 
Coulomb intra-grain interactions respectively. 
We fix the couplings as in~(\ref{single}) 
with the exception that 
\begin{eqnarray}
u_{j} = \Phi_{a} - \varepsilon_{j}/E_{D}, \mbox{when} \; j\in I_{a}.
\label{multi}
\end{eqnarray}
Now we impose
\begin{equation}
\Phi_{a+1}-\Phi_{a} \gg \max\limits_{j,k\in I_{a}}
\left\{\left(\varepsilon_{j}-\varepsilon_{k}\right)/E_D\right\}
\label{multi1}
\end{equation}
to make  the tunneling amplitude exponentially suppressed with 
the \emph{spatial} distance between the grains.
The pairing interaction is nearly uniform
for levels within $E_D$ in the same grain. 
The intra-grain Coulomb interaction is also nearly 
uniform $U_{jk}\simeq A$, while the inter-grain Coulomb interaction 
is modulated by the corresponding energy separation. 

\vspace{-0.2cm}
\bigskip

{\it Conclusions.}
We  found a class of  integrable 
Hamiltonians, which are a
generalization of the BCS Hamiltonian characterized by non-uniform 
coupling constants. To our knowledge this is the first exact
solution for non-uniform pairing interaction.
The strategy  we have  adopted consists in 
generalizing the procedure of Ref.~\cite{CAMBIAGGIO}, 
namely   constructing the Hamiltonian of the system 
in terms of anisotropic Gaudin Hamiltonians.
By means of  the integrability and the exact solvability of the latter
we  obtain the integrability and the exact solution of the 
model  Eq.~(\ref{our-model}), (\ref{coupling-constants}). 
In this sense, our 
procedure is close in spirit to the quantum 
inverse scattering method~\cite{KOREPIN-BOOK}. 
\\
The isotropic limit 
$q\to 0$ of the Gaudin Hamiltonians corresponds to uniform couplings. 
For arbitrary $A$, the Hamiltonian is the sum  of the BCS 
and the orthodox model. For $A=g$ the BCS Hamiltonian is obtained; 
the same isotropic limit of the exact solution  
Eqs.~(\ref{vector})-~(\ref{generalized-richardson}) 
coincides with the RS solution.   

This class of models might be useful for applications to the 
physics of metallic grains. 
The non-uniformity~\cite{NOTE5} 
of the coupling constants~(\ref{coupling-constants}) 
corresponds to include 
certain  ${\cal O}(\delta E^2/E_T)$ terms~\cite{AGAM1} 
in the Hamiltonian~(\ref{universal}). 
In fact, we  recover the 
``fluctuations'' of the Coulomb interaction 
of  the Ref.\cite{AGAM1} identifying   
${\delta U}_H\equiv U_{ij}-U_{ij'}$. 
The integrable model presented here 
might be applied as a starting point for suitable perturbation 
schemes leading to the explanation of the tunneling phenomena.

The present model can be applied to systems of 
few  interacting dots, since 
our  capacitive-like inter-grain  interaction does not decay 
with spatial distance.  
   
In a recent paper Ref.~\cite{DUKELSKY}  a non-uniform coupling 
for bosonic  systems was studied.
The Hamiltonian was  constructed from the bosonic analog of formulas 
(\ref{modified-crs}), (\ref{constants}), 
where  the $S^a$ 
are  generators of $su(1,1)$ (instead of $su(2)$). 
This algebraic difference does not affect the equations which  
$w_{ij}$, $v_{ij}$ have to fulfill to  ensure the commutativity 
of the (bosonic) $\tau_j$. 
The coupling constants of this bosonic model  
can be obtained in the isotropic limit of our Eqs.~(\ref{coupling-constants})
with $u_j\propto \varepsilon_j^d$ and $A=0$. This shows that 
the bosonic  Hamiltonian in Ref.~\cite{DUKELSKY} can be obtained by the  
limit $q \to 0$ of anisotropic 
$su(1,1)$ Gaudin models. Work is in progress along this direction.  

\vspace{-0.2cm}
\bigskip 
This line of research was suggested by G. Falci whose  
invaluable help is a pleasure to acknowledge.  
We thank R. Fazio for constant support, fruitful discussions and critical 
reading of the manuscript. We acknowledge R. W. Richardson for 
critical discussions on our work. We also thank A. Mastellone 
for useful discussions.


\end{multicols}


\begin{thebibliography}{100}
\bibitem{NOTE2} This factorization holds since (as in the BCS case)  
	the pairing interaction involves only doubly occupied levels,
	the magnetic term only singly occupied ones.
\bibitem{NOTE3} 
	This is due to the fact that the magnetic term commutes
	with the remaining ones in the Hamiltonian Eq.~(\ref{our-model})   
	as a consequence  of  the general property that the charge 
	realization of $su(2)$ is orthogonal to the  spin one.
\bibitem{NOTE4} The choice of $\varepsilon_1$, 
	 $\varepsilon_\Omega$ as cutoffs would imply 
	 dependency on the particular configuration of blocked levels.
\end{thebibliography}
\end{document}